\documentstyle[aps,epsf]{revtex}
\def\beq{\begin{equation}}
\def\eeq{\end{equation}}
\def\bea{\begin{eqnarray}}
\def\eea{\end{eqnarray}}
\def\eqref#1{Eq.~(\ref{eq:#1})}

\def\eqlab#1{\label{eq:#1}}
\def\vslash#1{\mbox{/\llap{#1} }}    

\begin{document}
\title{On electromagnetic off-shell effects in proton-proton brems\-strahlung}
\author{S. Kondratyuk, G. Martinus, O. Scholten}
\address{Kernfysisch Versneller Instituut, 9747 AA Groningen,
The Netherlands.}
%
\maketitle
\begin{abstract}

We study the influence of the off-shell structure of the
nucleon electromagnetic vertex on proton-proton
bremsstrahlung observables. Realistic choices for these off-shell
effects are found to have considerable influences on observables such
as cross sections and analyzing powers.
The rescattering contribution diminishes the
effects of off-shell modifications in negative-energy states.

\end{abstract}

\vspace{2cm}


\noindent
Corresponding author: \\
S. Kondratyuk, Kernfysisch Versneller Instituut \\
Zernikelaan 25, 9747 AA Groningen, the Netherlands  \\
e-mail:  KONDRAT@KVI.NL, \\
phone:  +31-(0)50-3633551, fax:    +31-(0)50-3634003 \\

\newpage

\section{Introduction}

Proton-proton bremsstrahlung is one of the simplest processes that can be
used to extract information on the off-shell nucleon-nucleon T-matrix from
experiments. The off-shell interaction lies at the basis of most
calculations on nuclear properties. To extract this interaction from the
data it is imperative that all competing effects are under control.
Elsewhere the effects of meson-exchange currents, isobar components and
relativity have been considered\cite{eden,mart,fearing}. In this letter we
want to address another competing mechanism, that of off-shell effects in
the electromagnetic nucleon vertex.

The problem of off-shell form factors in the $ \gamma NN $ vertex  has been
faced in different ways. In one
approach\cite{bincer,nyman70,nyman71,hare,davidson} one uses dispersion
integrals to relate the imaginary part of the vertex, arising from inelastic
processes, to the real part which now obtains an explicit dependence on the
invariant mass of the nucleon. An alternative
approach\cite{nauskoch,tiemtjon,dongmos,bosscherkoch} is to dress the vertex
explicitly with one-loop corrections. Also from Chiral Perturbation Theory
estimates can be obtained on the dependence of the off-shell form factor
near the on-shell point\cite{boskoch}. The different approaches give
considerable differences in their predictions. For this reason we propose a
parametrization of the off-shell form factor based on a low-energy behaviour.

We investigate effects of off-shell form factors on observables in
proton-proton bremsstrahlung using the fully relativistic framework
developed in ref\cite{mart}. The nucleon-nucleon T-matrix is calculated in
a Blankenbecler-Sugar approximation
by J. Fleischer and J.A. Tjon\cite{fltjon}. The
effects of the form factor on observables are discussed in terms of the
coefficients of our parametrization and the effects (for realistic
parameters) are found to be large.

\section{The electromagnetic off-shell form factors}

The most general structure of the nucleon-nucleon electromagnetic
vertex for the real photon reads\cite{bincer} \footnote{We follow the notations
of\cite{bjorkdrell} for the $\gamma$ - matrices and $\sigma_{\mu \nu}$}:
\begin{equation}
\Gamma_{\mu}(W,W^\prime)=e\gamma_{\mu}+i\kappa\!\sum_{rs=+,-} \!\Lambda^r(W)\,
\frac{\sigma_{\mu \nu} q^{\nu}}{2m}\, \Lambda^s(W^\prime)\,F^{rs}(W,W^\prime),
\eqlab{1}
\end{equation}
or equivalently
\beq
\Gamma_{\mu}(W,W^\prime)=e\gamma_{\mu}+i\kappa\!\sum_{kl=0,1} \!\vslash{p}^k\,
\frac{\sigma_{\mu \nu} q^{\nu}}{2m}\,
\vslash{p}'^l\,A_{kl}(W^2,W^{\prime 2}),
\eqlab{2}
\eeq
where $ e \,(m) $ is the proton charge (mass), $ \kappa $ is the anomalous
magnetic moment. In the formulas above the initial (final)
proton has momentum $ p^\prime \, (p),  q=p^\prime-p $ is photon momentum,
$ W=\sqrt{p^2}, W^\prime=\sqrt{p^{\prime 2}} $. The operators
$ \Lambda^{\pm}(W)=(\pm \vslash{p} + W)/(2W) $ project on
positive- and negative-energy states of the off-shell proton with
invariant mass $ W $. The functions
$ F^{rs}(W,W^\prime) $ or $A_{kl}(W^2,W^{\prime 2})$ are the
electromagnetic off-shell form factors of the nucleon.

Time reversal requires symmetries among the form factors,
 $ F^{rs}(W,W^\prime)= F^{sr}(W^\prime,W) $.
From the equivalence of \eqref{1} and \eqref{2} one deduces immediately
that
\beq
F^{rs} = A_{00}+r\, W\, A_{10}+s\, W^\prime A_{01}+r\,s\,W\,
W^\prime \, A_{11}, \mbox{\ \ with\ \ } r,s=+,- \ ,
\eeq
where $ A_{kl} $ are functions of $ W^2 $ and $ (W^\prime)^2 $.
$ F^{rs}$ thus obey the following symmetry relation
\beq
 F^{rs}(W,W^\prime)=F^{++}(r \,W,s \,W^\prime).
\eeq
Therefore, it suffies to know $ F^{++} $ only.


In our bremsstrahlung calculations we are interested in photon energies
up to 150 MeV.
Although this region extends beyond the applicability of
low-energy theorems, one may still use a low-energy expansion to
determine the most important aspects of off-shell form factors.
From a low-energy expansion of the bremsstrahlung matrix element
around the on-shell point
($ W=W^\prime=m $) one finds that the leading order contribution is
determined by two parameters:
the slope
\beq
k = \left. \frac{\partial F^{++}(W,W^\prime)}{\partial (W/m)}
\right|_{W=W^\prime=m}\,=
\left. \frac{\partial F^{++}(W,W^\prime)}{\partial (W^\prime/m)}
\right|_{W=W^\prime=m} \; ,
\eeq
and the value of
$ F^{+-}(W,W^\prime) $ and $ F^{-+}(W,W^\prime) $ at the on-shell point,
\beq
\kappa^{-} = F^{+-}(m,m)=F^{-+}(m,m) \;.
\eeq
%
Henceforth, we will use the values
for $k$ and $\kappa^-$ to compare the different
models for off-shell form factors. In all our calculations
the experimental value for the anomalous magnetic
moment of the proton will be used,
$\kappa=1.79$. We will also impose that Eq.(1) reproduces the
standard electromagnetic vertex
with all the particles on-shell and thus $F^{++}(m,m)=1$.

One approach to calculate off-shell form factors is through the application
of dispersion relations\cite{nyman70,nyman71,hare}. The real part of a form
factor is related to the integral of the imaginary part which in turn is
related to a total cross section by unitarity. This should give a model
independent estimate of the form factor. However, for the evaluation of the
integral the cross section at very high energies is required, which is
unknown. Related to this, the convergence of the dispersion integral is
unclear and several subtractions may be necessary. These unknowns are
reflected in a rather large range of values deduced for $k$ and $\kappa$.
Threshold dominance assumption was adopted in Refs.\cite{nyman70,hare} to
compute imaginary parts of the form factors. It led to the values  $
\kappa^- =24 $ and $ k=2.2 $ in \cite{nyman70}. In Ref. \cite{hare} instead
$ \kappa^- =14 $ was obtained. The value of $k$ is not given in this
reference, but seems considerably larger than the value quoted
in\cite{nyman70}. If more sutractions are needed to make the dispersion
integrals convergent, even these values are not predicted
anymore\cite{nyman71}.

A self-consistent model for the form factors near the on-shell point can be
obtained from next-to-leading-order Chiral Perturbation Theory. The
isovector part of $ k $ was calculated\cite{boskoch} to be equal to 2, with
the isoscalar part being negligible in the chiral limit. This approach does
not predict $ \kappa^{-} $.

Near the on-shell point one expects that off-shell corrections to the vertex
are calculable in a one loop model. H.W. Naus and J.H. Koch did calculations
with a dominant contribution from the one-pion loop \cite{nauskoch}, where
the pseudoscalar pion-nucleon coupling was used. Values $ k=-0.38 $ and $
\kappa^{-}=6.88 $ are inferred from this model. Another qualitative model
was developed in Ref.\cite{bosscherkoch}. There, the loop comprises nucleon
and a "dressing" scalar meson of mass 0.8 GeV. This model predicts values $
k=-0.24 $ and $ \kappa^{-}=1.53 $. A more realistic approach including the
vector dominance model was presented by P.C. Tiemeijer and J.A. Tjon
\cite{tiemtjon}. It leads to $ k=-0.23 $ for the pseudoscalar coupling and $
k=-0.32 $ for the pseudovector one. (The value of $ \kappa^{-} $ is not
presented in Ref. \cite{tiemtjon}.) In an analogous framework \cite{dongmos}
also the delta degree of freedom was included in the one-pion loop. This
model gives rise to values of approximately zero for both $ k $ and
$\kappa^{-} $.


The main conclusion we draw from this short overview of the results on the
form factors is that a consensus is lacking. For this reason we have
adopted an {\it ad hoc} parametrization of the form factors,
based on the low energy expansion, with $k$ and
$\kappa^-$ as parameters.
As was shown above,
it is enough to parametrize the
form factor $ F^{++}(W,W^\prime) $ only,
\begin{eqnarray}
F^{++}(W,W^\prime) & = & 1\,+\,\left(\frac{W-m}{m}\,\frac{a_1W+a_2W^\prime}{m}\,+
\frac{W^\prime-m}{m}\,\frac{a_1W^\prime+a_2W}{m}\right) \nonumber \\ \label{2}
 & & \,\,\times \exp\left({-\frac{(|W|^2-m^2)^2 +(|W^\prime|^2-m^2)^2}
{2 d m^4}}\right).
\end{eqnarray}
The constants $ a_1 $ and $ a_2 $ are easily expressed through the parameters
$ k $ and
$ \kappa^{-} $ discussed above: $ k=a_1+a_2,\, \kappa^{-}=1+2(a_1-a_2) $.
The exponential decreasing factor was included to
provide that the form factors do not grow like $W^2$ at infinity.
At the value of the constant $d$ used in the calculations, $d=0.5$,
this exponent is of a minor importance for the observables
(see discussion of the results below).

Since the values of $k$ and $\kappa^{-}$ differ
considerably for the various approaches to the form factors,
we performed calculations for two different choices
of the parameters (please note that $k=0$, $\kappa^-=1$
and $1/d=0$ corresponds to ignoring all off-shell dependence);

\begin{itemize}
\item[$ F^{-} $:]
 $ k=0 $, $ \kappa^{-}=0 $. For this choice of parameters the role of
the negative energy states is emphasized since near the on-shell point
$F^{++}$ is not affected but $ F^{+-}$ and $F^{-+} $ are. This
parametrization will be referred to as  "$ F^{-} $" and is close to
that following from the one-loop model of reference \cite{dongmos}.
\item[$ \partial F^{+} $:] $ k=2$, $\kappa^{-}=1 $. For this case the
anomalous magnetic moment for negative- and positive-energy states is
the same, but with a linear dependence on the invariant mass near the
on-shell point. This parametrization is thus referred to as "$ \partial
F^{+} $". The value $ k=2 $ is taken from the Chiral Perturbation
Theory\cite{boskoch}. Since this model does not
predict $ \kappa^{-} $, we put $ \kappa^{-}=1=F^{++}(m,m) $ in order to reduce
the number of independent parameters and to make it complementary to
the previous choice.
\end{itemize}

\section{Results and Discussion}

\begin{figure}
\epsfxsize 10 cm
\centerline{\epsffile[4 80 592 718]{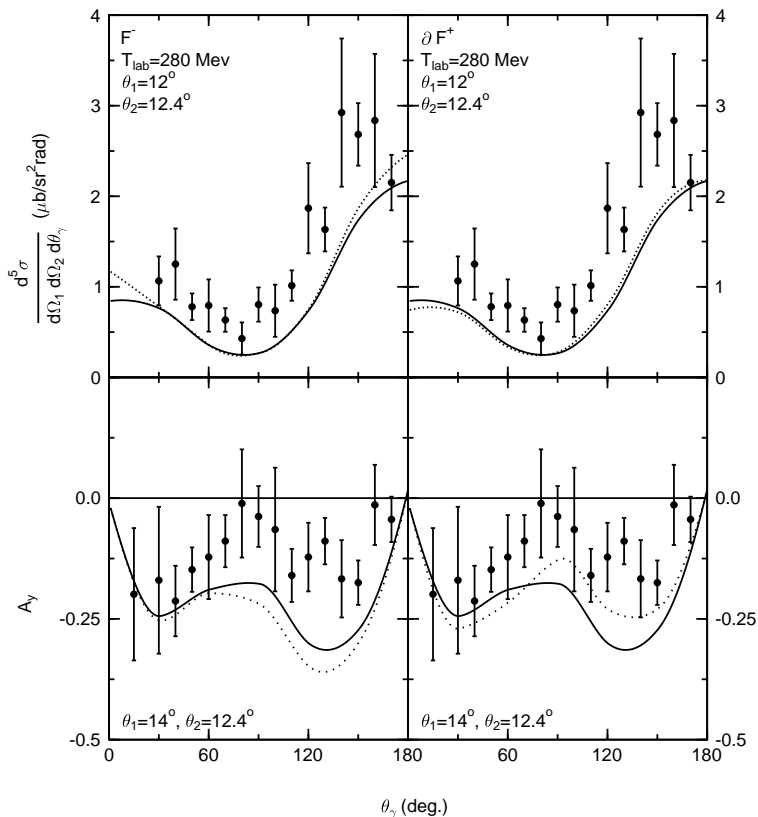}}
\caption[f1]{
Influence of the off-shell nucleon electromagnetic form factor
on the observables as function of the photon angle. $T_{lab}$ is the incoming
proton energy in the
laboratory frame, $\theta_{1}$ and $\theta_{2}$ are the outgoing
proton angles. Two upper panels show cross sections; two
lower show the analyzing powers. The solid (dotted) lines correspond to the
calculations without (with) including off-shell effects.
 Two left
panels correspond to the $ F^{-} $-case of parametrization; two right ones to
the $ \partial F^{+} $-case. The data are taken
from the TRIUMF experiments \cite{triumph}.
\label{fig1}}
\end{figure}

In the calculations we used the approach developed in Ref.\cite{mart},
based on a fully relativistic t-matrix of Ref.\cite{fltjon} in which
a complete account is given of relativistic effects. The rescattering
contribution was included in the calculations, which was shown to be crucial
for a correct treatment of negative-energy states. Corrections due to
meson-exchange currents and the isobar degree of freedom were not
included. These are not important for investigating
the influence of possible off-shell effects, but will be for a definitive
comparison with the data.

 Fig.1 shows that the off-shell effects change the cross section up to
$\approx$ 15\% for the $ \partial F^{+} $-case and $\approx$ 30\% for the
$F^{-}$-case. The effect is even larger for the analyzing power at a wide
range of photon angles. Corrections are seen of the order of $\approx$ 30\%
for the $\partial F^{+}$- and $\approx$ 25\% for the $F^{-}$-calculations.
We checked the effect of the parameter $d$ as well. Increasing $d$ has
negligible effects on the observables: less than 2\% for d varying from 0.5
to 1000. The latter value corresponds to the  practical absence of the
exponent in the parametrization of the form factors.

\begin{figure}
\epsfxsize 13 cm
\centerline{\epsffile[4 363 568 718]{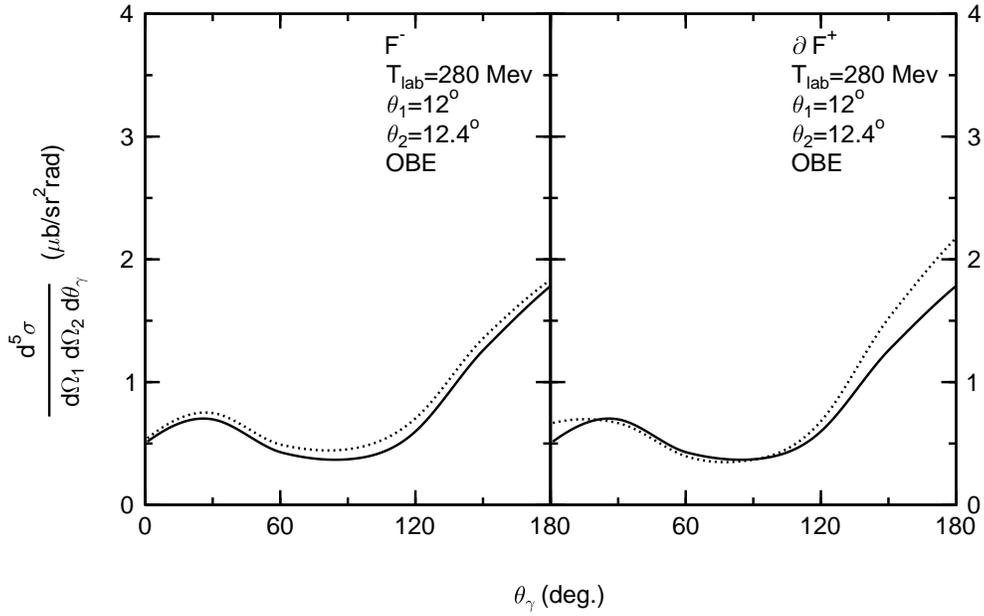}}
\caption[f2]{
The same as the upper panels of Fig.1, but calculated within the
one-boson-exchange model.
\label{fig2}}
\end{figure}

To investigate whether the off-shell effects found are due to the
one-boson-exchange kernel or they originate from the full complexity
of the t-matrix, we calculated the cross section with and without the
off-shell corrections within a one-boson-exchange model including $
\pi - ,\,\rho - ,\,\sigma - ,\,\omega - ,\,\eta - , $ and $ \delta - $
mesons. The coupling constants were taken the same as in the kernel of the
t-matrix. The result for the cross section is shown in Fig. 2 (the analyzing
power is zero in the one-boson-exchange model).
It is seen that the effects are of the same order of magnitude as
 in the full calculations. In detail there are however large differences,
while in the full calculations the effects of $\kappa^-$ are large, the
opposite is found in the obe-calculations.


To obtain a better understanding of these off-shell effects we have
investiated them in a low-energy approach.
The low-energy theorem for the bremsstrahlung processes \cite{low} states
that in the expansion of the cross section in the small energy $ q $
of the radiated photon,
\beq
\sigma=\frac{\sigma_0}{q}+\sigma_1+q\sigma_2+q^2\sigma_3+\cdots
\,, \eqlab{let}
\eeq
the terms $ \sigma_0 $ and $ \sigma_1 $ are uniquely determined. It means
that any off-shell effects may contribute
only to the coefficients $ \sigma_2 $,
$ \sigma_3 $, and terms of higher order in $ q $.
Within the one-boson-exchange model, we investigated analytically the
$q$-dependence of the difference $ \Delta\sigma $ between cross sections with
and without the off-shell corrections. In a low energy expansion of
$ \Delta\sigma $, we were retaining only leading orders in $ q $ and
$ p_{ext}/m $, $ p_{ext} $ being the general notation for the initial and
final proton three momenta. The
leading off-shell contribution was found to be
proportional to $ q $ for both the $ \partial F^{+} $
and $ F^{-} $ cases, i.e.
$ \Delta\sigma=q\sigma_2+q^2\sigma_3+...\ $, in accordance with \eqref{let}.
Let $\Delta\sigma(meson)$ denote a part of $ \Delta\sigma $
that comes from the
exchange of a particular $meson$. We found that for every meson,
$ \Delta\sigma(meson)$ is proportional to $q$.
More specifically, for the $ \partial F^{+} $-case of parametrization,
the ratio $\Delta\sigma(\rho)/\Delta\sigma(\omega) \approx
\Delta\sigma(\rho)/\Delta\sigma(\pi) \approx 2$.
For the $ F^{-} $-case, the $\omega$-meson dominates, with
$\Delta\sigma(\omega)/\Delta\sigma(\rho) \approx
\Delta\sigma(\omega)/\Delta\sigma(\pi) \geq 10$, depending on photon
angle.
At the same time, $\Delta\sigma(\omega)$ for the $ \partial F^{+} $
and $ F^{-} $-cases are of the same order of magnitude.
Since the contributions of the different mesons add constructively for the
$ \partial F^{+} $-case, after summing up all the mesons, $\Delta\sigma$ in the
$ \partial F^{+} $-case is much larger than $\Delta\sigma$ in the
$ F^{-} $-case as also seen in Fig.2.
These features can be explained
as follows. The $\omega NN$-coupling contains only the vector part,
and it can be
shown that the vector part of the meson-nucleon coupling leads to the same
$\Delta\sigma(meson)$ for both $ \partial F^{+} $ and $ F^{-} $.
The tensor part of
the $\rho NN$-coupling plays different roles for positive- and negative-
energies, which causes the apparent difference in contributions from
the $\rho$-meson for the calculations
with $ \partial F^{+} $- and $ F^{-} $-parametrizations.

\begin{figure}
\epsfxsize 10 cm
\centerline{\epsffile[17 94 553 718]{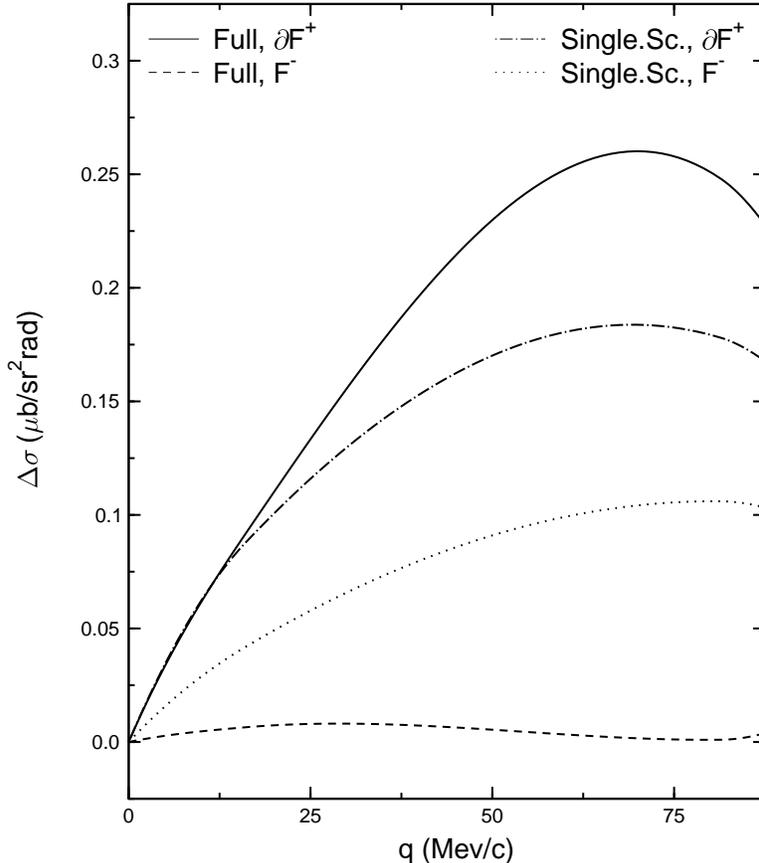}}
\caption[f3]{
The dependence on the photon energy of $\Delta\sigma$, the difference between
cross sections with and without the off-shell corrections to the
$\gamma NN$-vertex. All plots are obtained in the calculations with the full
t-matrix. The dash-dotted (dotted) line corresponds to the calculation where
only the single scattering diagrams were taken into account, for the case of
the $ \partial F^{+} $ ($ F^{-} $)-parametrization. The solid (dashed) line is
the calculation including the rescattering contribution, for the
$ \partial F^{+} $ ($ F^{-} $) case.
\label{fig3}}
\end{figure}

For a proper description of the off-shell effects in the electromagnetic
vertex it is necessary to include the rescattering contributions. This is
demonstrated in
Fig. 3, where the dependence is shown of $ \Delta\sigma $ on the photon energy
for the calculations based on the full t - matrix.
These calculations were done at $T_{lab}=280$ MeV and photon angle
$\theta_{\gamma}=5^{\circ}$. The photon energy was varied by changing
the (equal) angles of the outgoing protons from $33^{\circ}$ to $43^{\circ}$.
Comparing the full calculations for the $ F^{-} $ and $ \partial F^{+} $
cases,
we see that at low photon energies the off-shell effect coming form the negative-energy part of the
vertex is much smaller than that from the
positive-energy part, which is consistent with our
findings in the one-boson-exchange model and appears to be due to the
smallness of the $\omega$-exchange contribution for the full result. Eventhough in the present
calculations rescattering gives a similar suppresion of the effect in
negative-energy states as that of Ref.\cite{mart}, the situation is
qualitatively
different here. In ref.\cite{mart} the effect of negative-energy states in the
convection current was studied, where current concervation
was shown to put strong constraints on the leading order effect. In our
case the magnetic part of the vertex is addressed, where arguments based on
current concervation do not apply. Inspite of this, still the rescattering
contribution is responsible for a strong cancellation of the effect of
negative-energy states.
At the same time, the contributions to $ \Delta\sigma $
from negative- and positive-energies
are comparable with each other when the photon energy becomes larger than
100 MeV/c, which corresponds to the kinematics in figures 1 and 2.
It also follows from Fig. 3 that no contribution
of the off-shellness shows up either in the constant
$ \sigma_1 $ or in the singular term $ {\sigma_0}/{q} $ in Eq. (8), which
implies that the calculations fulfill the low-energy theorem.

We also did similar calculations for $T_{lab}=500$ MeV (not shown here).
We found that the off-shell effects on the observables are larger, but only
by at most a factor 1.5, compared to the calculations at $T_{lab}=280$ MeV.

In conclusion, we investigated the effects of the off-shell electromagnetic
structure
of the nucleon on observables in proton-proton bremsstrahlung. We used a
parametrization of the nucleon electromagnetic off-shell form factors based on
its general properties and with values consistent with model calculations.
We found a sizable influence of these off-shell effects
on the observables, in particular on the analyzing power.

\section{Acknowledgements}
We thank A.Yu. Korchin for valuable discussions concerning
off-shell effects at low energies. This work is a part of the
research of one of
the authors (S.K.) at Institute for Theoretical Physics, 252143 Kiev, Ukraine.


\end{document}